\documentclass[conference]{IEEEtran}
\IEEEoverridecommandlockouts

\usepackage{cite}
\usepackage{amsmath,amssymb,amsfonts}
\usepackage{algorithmic}
\usepackage{graphicx}
\usepackage{textcomp}
\usepackage{xcolor}
\usepackage{caption}
\usepackage{subcaption}

\def\BibTeX{{\rm B\kern-.05em{\sc i\kern-.025em b}\kern-.08em
    T\kern-.1667em\lower.7ex\hbox{E}\kern-.125emX}}
\begin{document}

\title{Performance Analysis of NOMA-Assisted Optical OFDM ISAC Systems with Clipping Distortion}
\author{\IEEEauthorblockN{Nam N. Luong, Chuyen T. Nguyen}
\IEEEauthorblockA{\textit{School of Electrical and Electronic Engineering} \\
\textit{Hanoi University of Science and Technology}\\
Hanoi, Vietnam \\
luongnamnd220503@gmail.com, chuyen.nguyenthanh@hust.edu.vn}
\and
\IEEEauthorblockN{Thanh V. Pham}
\IEEEauthorblockA{\textit{Dept. of Mathematical and Systems Engineering} \\
\textit{Shizuoka University}\\
Hamamatsu, Japan \\
pham.van.thanh@shizuoka.ac.jp}
}
\maketitle

\begin{abstract}
This paper studies the performance of optical orthogonal frequency-division multiplexing (OFDM)-based multi-user integrated sensing and communication (ISAC) systems employing non-orthogonal multiple access (NOMA). Due to their inherent high peak-to-average power ratio (PAPR), OFDM waveforms are clipped to fit the limited dynamic range of the optical transmitters (e.g., light-emitting diodes (LEDs)), resulting in clipping distortion.  To alleviate the impact of the distortion, we propose a novel transmitter architecture where the clipping processes are performed before NOMA superposition coding. We then analyze the performance of the proposed optical ISAC systems considering the effects of power allocation and clipping distortion. 
For the communication subsystem, we analyze the effect of NOMA on the achievable sum rate and bit error rate (BER). For the sensing subsystem, the root mean square error (RMSE) and Cramér-Rao bound (CRB) of estimating the transmission distance accuracy are obtained. Simulation results reveal that allocating more power to the strong user yields a higher sum rate, lower BER, and better sensing performance, whereas a more balanced power allocation among users results in degraded BER and sensing performance.
\end{abstract}

\begin{IEEEkeywords}Optical integrated sensing and communication (ISAC), optical orthogonal frequency-division multiplexing (OFDM), non-orthogonal multiple access (NOMA), clipping distortion, 
\end{IEEEkeywords}

\section{Introduction}
In recent years, integrated sensing and communication (ISAC) has witnessed significant advancements and attracted substantial research interest. ISAC has also been recognized by IMT-2030 as one of the six key usage scenarios for sixth-generation (6G) mobile networks \cite{IMT-2030}. In addition to radio-frequency (RF) ISAC systems, optical ISAC has emerged as a promising alternative solution owing to its unlicensed spectrum and immunity to electromagnetic interference \cite{wen2023optical}. 
Typically, optical ISAC employs laser sources for free-space optical (FSO) systems and light-emitting diodes (LEDs) for visible light communication (VLC) systems to simultaneously transmit data and sense the environmental information.

The performance of optical wireless communication systems, such as VLC, is generally constrained by the low modulation bandwidth of the LEDs (i.e., typically a few MHz). To enhance the spectral efficiency, optical orthogonal frequency-division multiplexing (O-OFDM) is commonly employed. However, it is well-known that OFDM waveforms generally exhibit a high peak-to-average power ratio (PAPR) due to the combination of independently phased subcarriers. To accommodate the limited dynamic range of LEDs, OFDM waveforms are usually subjected to clipping, which introduces clipping distortion and can significantly degrade the overall system performance \cite{Thanhss}.

The impact of clipping distortion in optical OFDM-based ISAC systems has been investigated in several recent works. In \cite{PADCO}, the authors investigated an ISAC system employing direct current-biased optical (DCO)-OFDM and light detection and ranging (LiDAR), where clipping distortion was considered and power allocation for the DC bias was optimized. In \cite{10960484}, the impact of clipping distortion on enhanced asymmetrically clipped optical OFDM (EADO-OFDM) was analyzed, and a power allocation scheme between DCO-OFDM and ACO-OFDM components was proposed.
For VLC systems, positioning is typically achieved using the integrated positioning and communication (IPAC) technique \cite{IPAC}, \cite{localizationVLC}, where the receiver estimates its position based on the received signal strength (RSS) or by employing ISAC. However, using ISAC in VLC systems in previous works often requires the use of a pinhole camera or additional optical components at the transmitter instead of processing the reflected signal as RF ISAC, which increases system complexity and cost \cite{pinhole}. The study in \cite{retroreflective2024} experimentally demonstrated a VLC-ISAC system employing a corner cube reflector (CCR) to exploit the reflected signal for sensing. The authors also derived an analytical expression for the sensing channel gain. 
However, their analysis was confined to a single point-to-point communication and sensing link, without addressing multiuser scenarios that are essential for practical VLC systems. To support multiple users with a single LED, non-orthogonal multiple access (NOMA) has been widely adopted in VLC systems. Nevertheless, existing NOMA-based studies \cite{effectclipping, electronics9101743, GEBEYEHU2020e03363} mainly focused on analyzing the impact of clipping noise when two OFDM signals are superimposed prior to clipping, while overlooking integrated sensing functionality and the joint optimization of communication and sensing performance.

To overcome the limitations of the aforementioned studies, in this work, we propose a DCO-OFDM-based optical ISAC system that employs NOMA to simultaneously serve multiple users. Unlike the approach in \cite{effectclipping}, where clipping is done after NOMA superposition, our system integrates NOMA and DCO-OFDM in such a way that multiple OFDM signals are individually clipped before the NOMA superposition. By doing so, the proposed design can effectively mitigate clipping distortion propagation, and simulation results show that the proposed approach achieves a lower PAPR compared with the study in \cite{effectclipping}. Based on this configuration, we investigate the joint impact of NOMA and clipping distortion on both communication and sensing performance. Specifically, for the communication channel, we evaluate the bit error rate (BER) and the achievable users' sum rate performance. For the sensing functionality,  we derive the root mean square error (RMSE) and the Cramér-Rao bound (CRB) of link distance estimation to quantify sensing accuracy.

The remainder of the paper is organized as follows. Section II describes the system and signal models. Performances of the communication and sensing subsystems are analyzed in Section III. Simulation results and related discussions are provided in Section IV. Finally, we conclude the paper in Section V. 
\begin{figure}[!t]
    \centering
    \includegraphics[scale = 0.35]{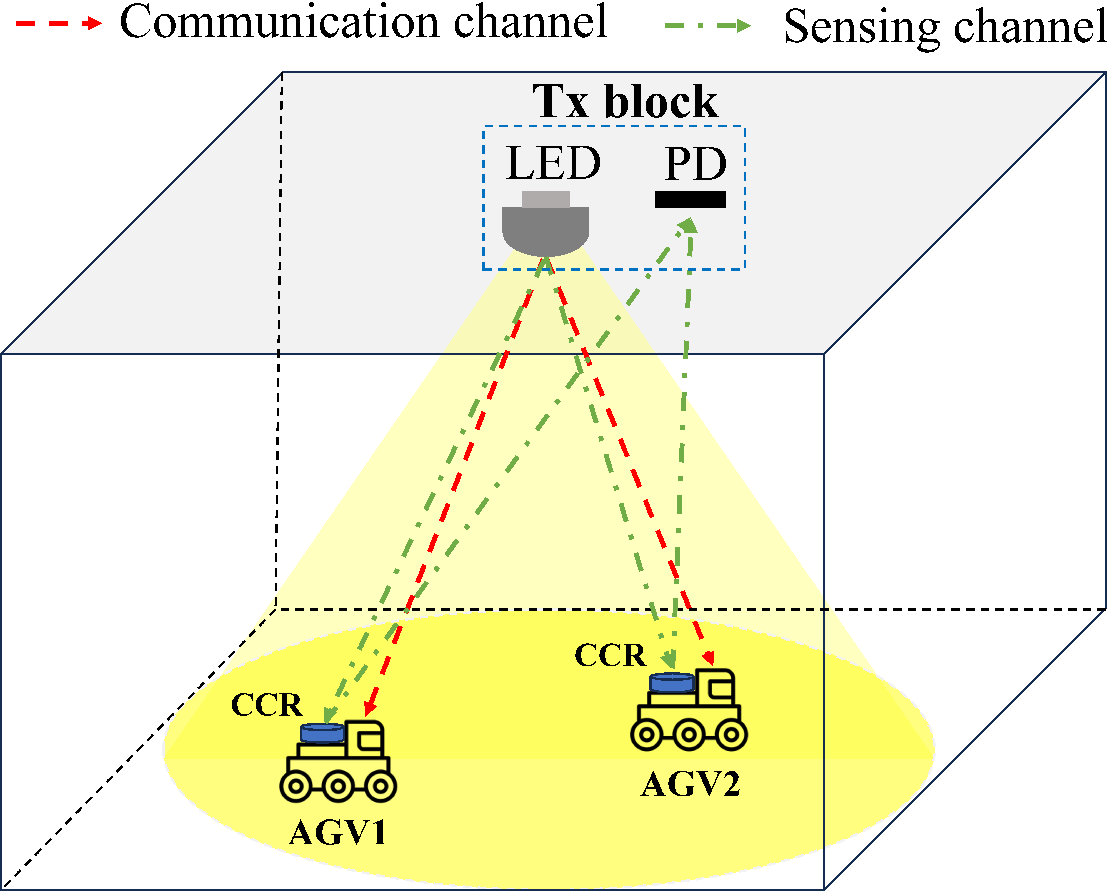}
    \caption{An example of NOMA-assisted multi-user optical ISAC system.}
    \label{fig:system_config}
\end{figure}
\section{SYSTEM MODEL}
\begin{figure*}[!t]
    \centering
    \includegraphics[scale = 0.525]{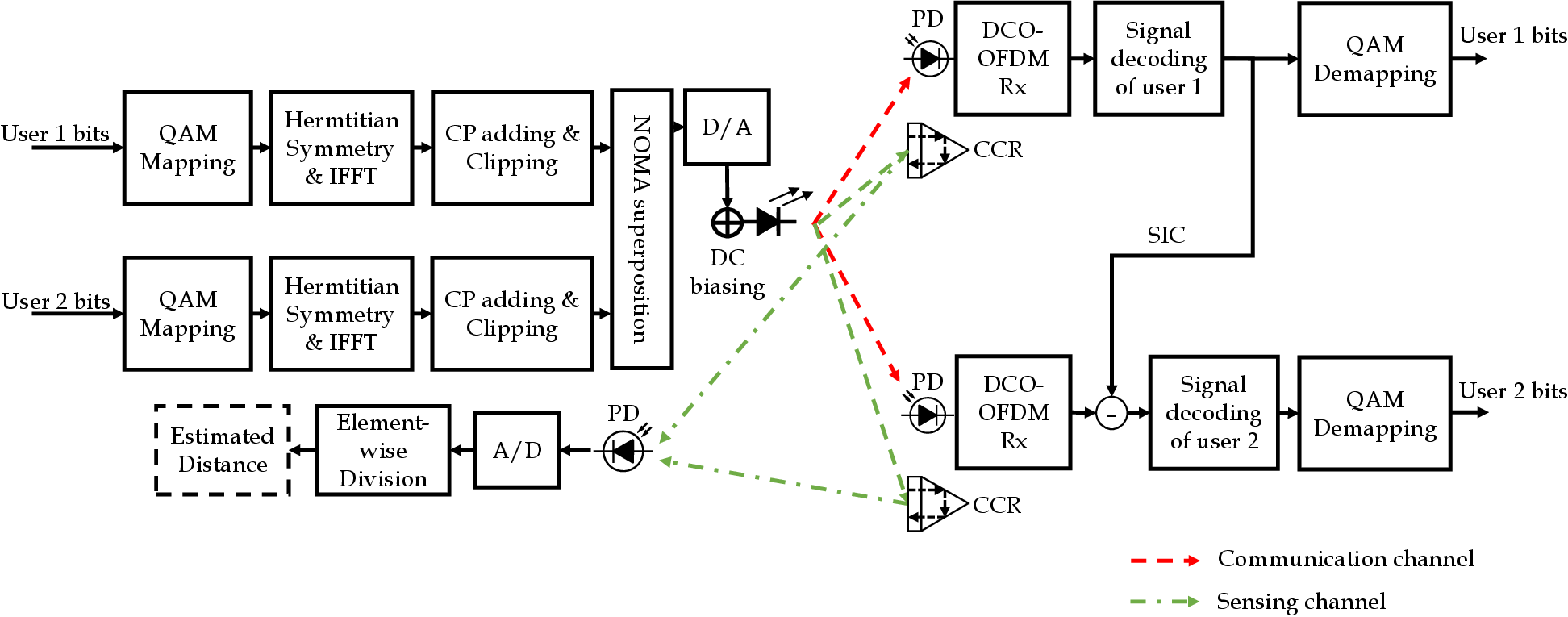}
    \caption{Schematic diagram of the proposed NOMA-assisted DCO-OFDM optical ISAC system.}
    \label{fig:system_model}
\end{figure*}
We consider an optical ISAC system employing DCO-OFDM\footnote{The analysis in the paper can be readily applied to other optical OFDM variants, such as ACO-OFDM. For conciseness, we consider DCO-OFDM in this work.} and NOMA for indoor scenarios, as illustrated in Fig.~\ref{fig:system_config}, where a single LED luminaire serves as the transmitter and two automated guided vehicles (AGVs) act as the users. For the sensing functionality, each AGV is equipped with a CCR to reflect the incident optical signal back toward the transmitter block which is equipped with a photodiode (PD). 
\subsection{DCO-OFDM Signal Model and Clipping Distortion}
DCO-OFDM is an OFDM variant specifically developed for intensity modulation/direct detection (IM/DD)-based optical wireless communication systems, in which the transmitted time-domain waveform is required to be real-valued and non-negative. 
In the considered multi-user DCO-OFDM system, each user’s input bit sequence is mapped onto an $M$-ary QAM constellation, resulting in frequency-domain symbols $X_i(k,l)$, where $X_i(k,l)$ denotes the symbol transmitted on the $k$-th subcarrier of the $l$-th OFDM symbol for user $i \in \{1,2\}$. Prior to performing the inverse fast Fourier transform (IFFT), to ensure that the resulting time-domain waveform is real-valued, the frequency-domain symbols are organized to satisfy Hermitian symmetry, i.e., $X_i(k,l)=X_i^*(N-k,l)$ and $X_i(0,l) = X_i(N/2,l) = 0$. The corresponding time-domain OFDM signal is given by
\begin{align}
x_i[n]=\frac{1}{\sqrt{N}}\sum_{l=0}^{L-1}\bigg\{\sum_{k=0}^{N-1}\left[X_i(k,l)e^{j2\pi k\Delta f(n-lT_0)}\right]\notag\\\operatorname{rect}\bigg(\frac{n-lT_0}{T_0}\bigg)\bigg\},
\label{OFDM_TD}
\end{align}
where $L$, $N$, $\Delta f$, and $\operatorname{rect}(\cdot)$ denote the number of OFDM symbols in an OFDM frame, the number of subcarriers, the subcarrier spacing, and the rectangular function, respectively. The total OFDM symbol duration $T_0$ consists of the duration of an elementary OFDM symbol $T_e$ and the guard interval $T_g$ added to mitigate inter-symbol interference, i.e., $T_o=T_e+T_g$. Due to the band-limited nature of the VLC channel, the OFDM signal is restricted to a fixed bandwidth $B = N\Delta f$ with $\Delta f = 1/T_e$. A cyclic prefix (CP) is subsequently appended to the signal to mitigate inter-symbol interference (ISI) and maintain subcarrier orthogonality.

To ensure the signal waveform remains within the dynamic range of the LED, the time-domain signal $x_i[n]$ is first clipped in the digital domain. Let $I_{\text{min}}$ and $I_{\text{max}}$ be the lower and upper limits of the linear dynamic range of the LED. Before NOMA superposition coding and DC biasing,  $x_i[n]$ is clipped as follows
\begin{align}
\tilde{x}_i[n] = \begin{cases}I_{\text{min}}-I_{\text{DC}},\ &\text{if}\ x_i[n] < I_{\text{min}},\\x_i[n]-I_{\text{DC}},\ &\text{if}\ I_{\text{min}}\leq x_i[n] \leq I_{\text{max}},\\I_{\text{max}}-I_{\text{DC}}, &\text{if}\ x_i[n]>I_{\text{max}},\end{cases}
\end{align}
where $I_{\text{DC}}$ denotes the DC bias. 

According to the Bussgang theorem, the clipped signal $\tilde{x}_i$ can be represented as
\begin{align}
\tilde{x}_i[n] = R_ix_i[n]+z_{\text{clip},i}[n],
\end{align}
where $R_i$ is the attenuation factor and $z_{\text{clip},i}[n]$ is the additive clipping noise. Let $\sigma^2_{x_i}$ be the average signal power of $x_i$, which, in the case of DCO-OFDM, is $\sigma^2_{x_i}=\bigg(1-\frac{2}{N}\bigg)\mathbb{E}[X_i^2(k, l)]$
and denote $\alpha_i=\frac{I_{\text{min}}-I_{\text{DC}}}{\sigma_{x_i}}$ and $\beta_i=\frac{I_{\text{max}}-I_{\text{DC}}}{\sigma_{x_i}}$, the attenuation factor is given by $R_i=Q(\alpha_i)-Q(\beta_i)$ \cite{clipping2012}, where $Q(t)=\frac{1}{\sqrt{2\pi}}\int_t^{\infty}\exp(\frac{-x^2}{2})\text{d}x$ is the Q-function. The clipping noise $z_{\text{clip},i}$ can be well approximated by a zero-mean Gaussian distribution whose variance is given by
 \cite{clipping2012}
\begin{align}
\sigma_{\text{clip}, i}^2=&\sigma_{x_i}^2\big(R_i+\alpha_i\phi(\alpha_i)-\beta_i\phi(\beta_i)+\alpha_i^2(1-Q(\alpha_i))\notag\\ &+\beta_i^2Q(\beta_i)- (\phi(\alpha_i)-\phi(\beta_i)+(1-Q(\alpha_i))\alpha_i \notag \\ & +Q(\beta_i)\beta_i)^2-R_i^2\big),
\end{align}
with $\phi(t)=\frac{1}{\sqrt{2\pi}}\exp(\frac{-t^2}{2})$. To facilitate the analysis, we assume identical average signal power for both users, resulting in 
$R_i$'s = $R$, $\alpha_i$'s = $\alpha$, and $\beta_i$'s =  $\beta$.
\subsection{NOMA Signal Model}
To enable simultaneous communication and sensing for multiple users using a single LED, we employ NOMA, which has been proven to offer higher spectral efficiency than orthogonal multiple access (OMA) schemes (e.g., time-division multiple access (TDMA)). After the clipping, the clipped OFDM signals for the two users are superimposed in the power domain according to the NOMA principle as illustrated in Fig.~\ref{fig:system_model}, resulting in    
\begin{align}
x_{\text{noma}}[n] &= \sqrt{\gamma}\tilde{x}_1[n] + \sqrt{1- \gamma}\tilde{x}_2[n] \notag \\
&= \sqrt{\gamma}(R x_1[n] + z_{\text{clip}, 1}[n]) \notag \\
&\quad + \sqrt{(1 - \gamma)}(Rx_2[n] + z_{\text{clip}, 2}[n]) \notag \\
&= \sqrt{\gamma}R x_1[n] + \sqrt{(1 - \gamma)}R x_2[n] + \tilde{z}_{\text{clip}}[n],
\end{align}
where $\gamma$ and $1 - \gamma$ ($\gamma > 0$) are the power allocation (PA) coefficients for the signals of the first and second users, respectively. Note that $\tilde{z}_{\text{clip}}[n] = \sqrt{\gamma}z_{\text{clip}, 1}[n] + \sqrt{(1 - \gamma)} z_{\text{clip}, 2}[n]$ follows a Gaussian distribution with zero-mean and variance $\sigma^2_{\text{clip}}$ since $z_{\text{clip},1}[n] \sim \mathcal{N}(0, \sigma^2_{\text{clip}})$ and $z_{\text{clip},2}[n] \sim \mathcal{N}(0, \sigma^2_{\text{clip}})$.

After the NOMA superposition, the digital signal $x_{\text{noma}}[n]$ is transformed into the analog domain, followed by DC biasing. The resulting waveform is then used as the input to the LED, which produces the output optical signal expressed by
\begin{align}
x_t[n] = \eta (x_{\text{noma}}[n]+I_{\text{DC}}),
\end{align}
where $\eta$ is the electrical-to-optical conversion factor. 
\subsection{Discussions on PAPR}
Since clipping distortion is directly proportional to the PAPR statistic of the OFDM waveform, we discuss in this section the PAPR in our proposed system with that in \cite{effectclipping}. The PAPR of the time-domain OFDM signal given in \eqref{OFDM_TD} is defined by 
\begin{align}
    \text{PAPR}\left\{x_i[n]\right\} = \frac{\underset{0 \leq n \leq N - 1}{\text{max}} \left|x_i[n]\right|^2}{\mathbb{E}\left[|x_i[n]|^2\right]}. 
    \label{PAPR_definition}
\end{align}
To assess the severity of PAPR, the complementary cumulative distribution function (CCDF) is often employed, 
which represents the probability that the PAPR \eqref{PAPR_definition} exceeds a given threshold $\text{PAPR}_0$, i.e., $\text{CCDF} = \text{Pr}(\text{PAPR} \geq \text{PAPR}_0)$.

\begin{figure}[!h]
    \centering
    \includegraphics[scale = 0.45]{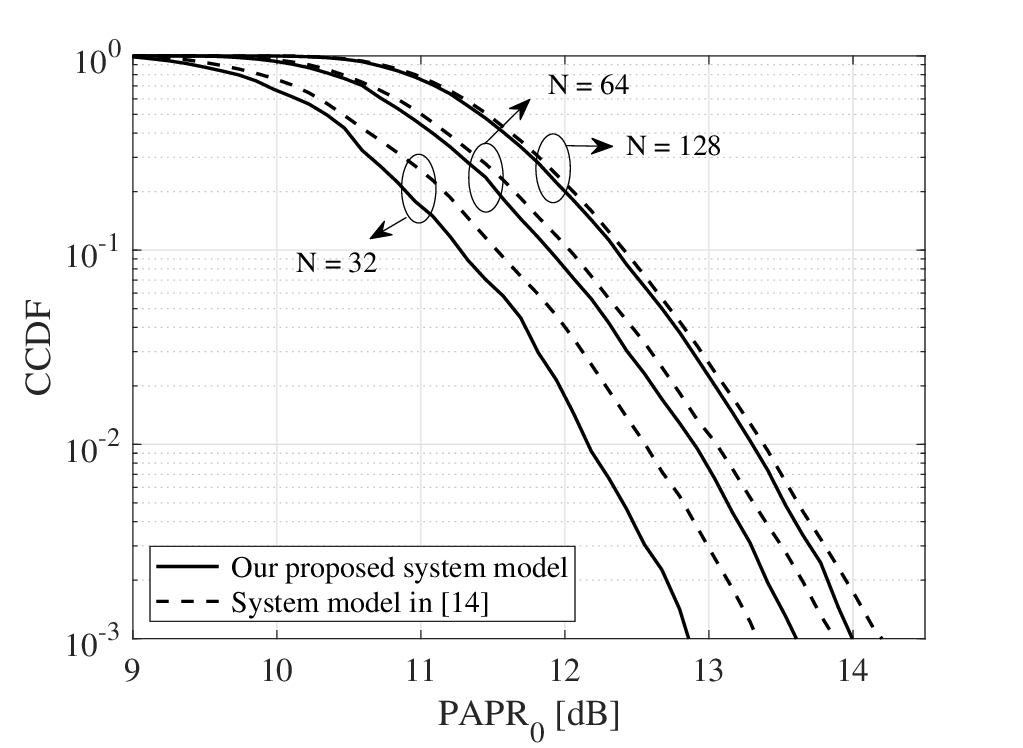}
    \caption{PAPR comparison between two system models.}
    \label{fig:papr}
\end{figure}
As shown in Fig.~\ref{fig:papr}, the PAPR of the signal obtained after NOMA superposition of multiple DCO-OFDM waveforms, as in \cite{effectclipping}, is clearly higher than that of each individual DCO-OFDM signal, as in our proposal, particularly when the number of subcarriers $N$ is small. This behavior arises from the heavier amplitude distribution tails in low-subcarrier cases, where the distribution is less sharply concentrated compared with that for a larger number of subcarriers. Consequently, the likelihood of peak summation increases, resulting in higher PAPR and more severe clipping distortion. As the number of subcarriers increases, this difference gradually diminishes. Since the effectiveness of NOMA also depends on the power allocation between users, we further analyze the impact of the PA coefficient $\gamma$ on both communication and sensing performance in the following section.
\section{Performance Analysis}
\subsection{Communication Subsystem}
Given that the line-of-sight (LoS) component of the VLC channel typically dominates the non-line-of-sight (NLoS) counterpart, this study considers only the LoS path to facilitate analytical tractability \cite{Pham2017}. For the $i$-th user, the channel gain of the communication subsystem is given by
\begin{align}
h_{\text{com},i}=\begin{cases}\frac{(m+1)\mathcal{A}_{\text{com}}\rho_{\text{com}}}{2\pi d_i^2}\cos^m(\varphi_i)\kappa(\theta_i)\cos(\theta_i),\ \\\text{for}\ 0\leq\theta_i\leq \Phi_{c}\\0, \ \text{otherwise,}\end{cases}
\end{align}
where $m=-\frac{1}{\log_2(\cos(\Psi))}$ is the Lambertian emission order with $\Psi$ being the semi-angle at half power of the LED.  $\mathcal{A}_{\text{com}}$ and $\rho_{\text{com}}$ are the active area and responsivity of the PD at the AGV, respectively. Also, $d_i$, $\varphi_i$, and $\theta_i$ are the link distance, the irradiance angle, and the incident angle of the optical signal.  $\kappa(\theta_i)=\frac{T_s(\theta_i)n^2}{\sin^2(\Phi_{c})}$, where $T_s(\theta_i)$ is the gain of optical filter, $n$ and $\Phi_{c}$ are the refractive index and field of view (FOV) of the receiver's optical lens.

The received time-domain electrical signal is given by
\begin{align}
y_{\text{com},i}[n]=h_{\text{com},i}x_t[n]+z_{\text{com},i}[n],
\end{align}
where $z_{\text{com},i}[n] \sim \mathcal{N}(0, BN_0)$ denotes the additive white Gaussian noise (AWGN) with zero mean and variance $BN_0$ where $N_0$ is the noise power spectral density. Note that the DC bias is filtered out since it does not carry information. After performing the analog-to-digital conversion and removal of CP, the time-domain signal $y_i[n]$ is transformed to the frequency-domain signal using FFT as 
\begin{align}
Y_{\text{com},i}(k,l)&=\eta h_{\text{com},i}(\sqrt{\gamma_1}RX_1(k,l)+\sqrt{\gamma_2}RX_2(k,l)\notag\\&+\tilde{Z}_{\text{clip}}(k))+Z_{\text{com},i}(k).
\end{align}
Without loss of generality, assume that $d_1<d_2$. As a result, User~1 has a better channel gain and thus is referred to as the strong user. Following the NOMA decoding principle, the signal of user 1 is decoded first by considering the signal of User~2 as noise. After that, successive interference cancellation (SIC) is applied to decode the signal of User~2. 
Since the power allocation scheme directly influences the clipping noise level, it consequently affects the achievable NOMA sum-rate for the two users. To quantify the impact of the PA coefficient, the rate of each user is formulated as \cite{effectclipping}
\begin{align}
\mathcal{R}_1&=\frac{1}{2}\log_2\bigg(1+\frac{h_{\text{com},1}^2R^2\gamma\sigma^2_{x}}{h_{\text{com},1}^2\sigma^2_{\text{clip}}+BN_0}\bigg),
\end{align}
and 
\begin{align}
\mathcal{R}_2&=\frac{1}{2}\log_2\bigg(1+\frac{h_{\text{com},2}^2R^2(1-\gamma)\sigma^2_{x}}{h_{\text{com},2}^2\sigma^2_{\text{clip}}+h_{\text{com},2}^2R^2\gamma\sigma^2_{x}+BN_0}\bigg).
\end{align}

\subsection{Sensing Subsystem}
VLC channels using CCRs for the sensing function are modeled based on two optical source models, namely: the point source and the area source \cite{retroreflective2024}. In this study, we adopt the channel gain of the area source model, as it can simultaneously support illumination, which is expressed as follows
\begin{align}
h_{\text{sen},i}=\begin{cases}\frac{(m+1)\mathcal{A}_{\text{sen}}\rho_{\text{sen}}\varsigma}{8\pi d_{i}^2}\cos^{m+1}(\varphi_i)\kappa(\varphi_i)\cos(\theta_i)\xi(\varphi_i,\theta_i),\\ \text{for}\ 0\leq\theta_i\leq \Phi_{\text{CCR}}\ \text{and} \ 0\leq\varphi_i \leq \Phi_{c}\\0, \ \text{otherwise,} \end{cases}
\end{align}
where $\mathcal{A}_{\text{sen}}$ and $\rho_{\text{sen}}$ denote the active area and responsivity of the PD at the transmitter, respectively. 
The parameters $\varsigma$ and $\Phi_{\text{CCR}}$ represent the reflectance and half-angle FOV of the CCR at the receiver. 
The term $\xi(\varphi_i, \theta_i)$ denotes the effective reflection ratio of the CCR with respect to the LED, which is defined as
\begin{align}
\xi(\varphi_i, \theta_i) = 
\frac{\mathcal{A}_{\text{eff}}(\varphi_i, \theta_i)}
{\mathcal{A}_{\text{LED}} - \mathcal{A}_{\text{sen}}},
\end{align}
where $\mathcal{A}_{\text{LED}}$ is the LED area, and $\mathcal{A}_{\text{eff}}(\varphi_i, \theta_i)$ represents the effective reflecting area of the CCR with respect to the LED, which can be obtained from \cite{RVLC2023} as
\begin{align}
\mathcal{A}_{\text{eff}}&(\varphi_i,\theta_i)=\frac{2r^2}{\cos(\varphi_i)}\bigg(\arccos\bigg(\frac{2(L_t+L_s)\tan(\theta_i)}{r}\bigg)\notag\\&-\frac{2(L_t+L_s)\tan(\theta_i)}{r}\sqrt{r^2-4(L_t+L_s)^2\tan^2(\theta_i)}\bigg)\notag\\&-\frac{A_{\text{sen}}}{\cos{(\varphi_i)}},
\end{align}
where $r=D_{\text{CCR}}+\frac{\sqrt{\mathcal{A}_{\text{sen}
}}}{2}$, $D_{\text{CCR}}, L_s, L_t$ are the diameter, the recessed length, and the length of CCR, respectively.

Similarly to the communication subsystem, the received sensing time-domain signal from the $i$-th user to the PD at the transmitter is given by
\begin{align}
y_{\text{sen},i}[n]=h_{\text{sen},i}x_t[n-\tau_{0,i}]+z_{\text{sen},i}[n],
\end{align}
where $z_{\text{sen},i}[n] \sim \mathcal{N}(0, BN_0)$ denotes the AWGN in the transmitter, $\tau_{0,i}$ is the true delay of the time-domain sensing signal of $i$-th user. 

After obtaining the electrical signal, element-wise division is performed with respect to the transmitted signal \cite{sturm2009novel}. The estimated delay of the received signal $\tilde{\tau}_i$ is then estimated by taking the IFFT along the subcarrier axis of the ratio between $Y_{\text{sen},i}(k,l)$ and $X_{\text{noma}}(k,l)$ which can be represented as follows
\begin{align}
\frac{Y_{\text{sen},i}(k,l)}{X_{\text{noma}}(k,l)}&=\frac{\eta h_{\text{sen},i}(X_{\text{noma}}(k,l))e^{-j2\pi k\Delta f\tau_{0,i}}+Z_{\text{sen},i}(k)}{X_{\text{noma}}(k,l)}\notag\\&=\eta h_{\text{sen},i}e^{-j2\pi k\Delta f\tau_{0,i}}+W_i(k),
\end{align}
where the frequency-domain noise
\begin{align}
W_i(k)\sim\mathcal{CN}\bigg(0, \frac{BN_0}{R^2\sigma^2_x+\sigma^2_{\text{clip}}}\bigg)
\end{align}
follows a complex Gaussian distribution with zero mean and variance $\frac{BN_0}{R^2\sigma^2_x+\sigma^2_{\text{clip}}}$. 
The estimated delay can be obtained as follows
\begin{align}
\tilde{\tau}_i= \underset{\tau}{\operatorname{argmax}}\bigg\{\frac{1}{\sqrt{N}}\sum_{k=0}^{N-1}\bigg(\frac{Y_{\text{sen},i}(k,l)}{X_{\text{noma}}(k,l)}\bigg)e^{\frac{j2\pi kn}{N}}\bigg\},
\end{align}
which is then used to give an estimated distance  $\tilde{d}_i=\frac{c\tilde{\tau}_i}{2}$ with $c$ being the speed of light. To quantitatively assess the accuracy of the distance estimation, we employ the root mean square error (RMSE), which measures the average deviation between the estimated distances and the true distances over multiple OFDM frames
\begin{align}
\text{RMSE}_i=\sqrt{\frac{1}{F}\sum_{f=0}^{F-1}\big(\tilde{d}_i[f]-d_i[f]\big)^2},
\end{align}
where $F$ denotes the number of OFDM frames. Furthermore, to evaluate the estimation performance, we derive the Cramér–Rao bound (CRB) for distance estimation following the approach in \cite{PADCO}, which is given by
\begin{align}
\operatorname{var}\{\tilde{d}_i\}\geq \frac{3c^2}{8\pi^2(\eta h_{\text{sen},i}\Delta f)^2\Lambda(N(N+1)(2N+1))}.
\end{align}
Here, $\Lambda=\frac{R^2\sigma_x^2+\sigma^2_{\text{clip}}}{BN_o}$ represents the ratio between the total power of the transmitted signal, including the clipping noise, and the noise power at the transmitter. In this work, the clipping noise is treated as a component that carries information about the delay, since it is part of the transmitted signal. Consequently, the derived CRB expression differs from that in \cite{PADCO}, where the clipping noise is assumed to carry no information about the delay. Furthermore, as the CRB represents the lower bound on the mean squared error (MSE) for any unbiased estimator, its square root should be taken when comparing with the RMSE.
\section{Simulation Results and Discussions}
In this section, the communication and sensing performances of the proposed system are evaluated through numerical simulations. Without otherwise noted, the simulation parameters are summarized in Table~I. For location specification of the transmitter and users, a Cartesian coordinate system is adopted, with the origin at the center of the floor.
\begin{table}
\vspace{2mm}
\centering
\caption{Simulation Parameters.}
\label{tab:parameters}
\begin{tabular}{l l}
\hline\hline
\textbf{System parameter} & \textbf{Value} \\ 
\hline\hline
$I_{\text{min}},\ I_{\text{max}},\ I_{\text{DC}}$ & 100~mA, 1000~mA, 500~mA\\ \hline
OFDM symbols in each frame, $L$ & 32 \\ \hline
OFDM frames, $F$ & 1000 \\ \hline
QAM order, $M$ & 4 \\ \hline
Subcarriers in each OFDM symbol, $N$ & 256 \\ \hline
OFDM bandwidth, $B$ & 20~MHz \\ \hline
Duration of a guard interval, $T_g$ & $\frac{T_e}{5}$ \\ \hline
Noise power spectral density, $N_0$ & $10^{-22}$~A$^2$/Hz \cite{Grubor2008}\\
\hline\hline
\textbf{Channel parameter} & \textbf{Value} \\ 
\hline\hline
Room dimension & \\ 
Length $\times$ Width $\times$ Height & 10 m $\times$ 10 m $\times$ 7 m  \\ \hline
LED location & $[0, 0, 7]$ \\ \hline
User 1 location & $[0, 0, 1]$ \\ \hline
User 2 location & $[2, 3, 1]$ \\ \hline
$\mathcal{A}_{\text{com}},\mathcal{A}_{\text{sen}}$ & 10~$\text{mm}^2$ \\ \hline
$\rho_{\text{com}},\rho_{\text{sen}}$ & 0.25 ~A/W \cite{retroreflective2024} \\ \hline
$\varsigma, n, T_s$ & 0.92, 1.51, 0.9 \\ \hline
$\mathcal{A}_{\text{LED}}$ & 50~$\text{cm}^2$ \\ \hline
$\Phi_c, \Psi$ & $60^{\circ}$ \\ \hline
$D_{\text{CCR}}, L_s, L_t$ & 50~mm, 6.3~mm, 35.7~mm\\ \hline
The speed of light, $c$ & $3\times10^8$ ~m/s\\ \hline
\hline
\label{fig:Table 1}
\end{tabular}
\end{table}

We first perform system simulations to evaluate the BER performance of the proposed system under various power allocation factors, i.e., $\gamma = 0.5$, $0.6$, $0.7$, and $0.8$. 
The BER is plotted as a function of the average transmit bit energy-to-noise power spectral density ratio, computed as transmit $
\frac{E_b}{N_0} = \frac{\sigma_x^2}{\log_2(M)\frac{N-2}{N}BN_0}.
$
Simulation results show that as $\frac{E_b}{N_0}$ increases, a more balanced power allocation between the two users (i.e., $\gamma$ approaching 0.5) leads to higher BER. Furthermore, when $\frac{E_b}{N_0}$ exceeds a certain threshold, the BER performance saturates. For instance, at $\gamma = 0.7$, the BERs of both users converge to approximately $10^{-2}$ once $\frac{E_b}{N_0}$ surpasses $100$ dB. This saturation occurs because, in the high $\frac{E_b}{N_0}$ region, clipping distortion becomes the dominant impairment, and further increases in $\frac{E_b}{N_0}$ no longer improve performance. 

\begin{figure}[t]
    \centering
    \includegraphics[scale = 0.45]{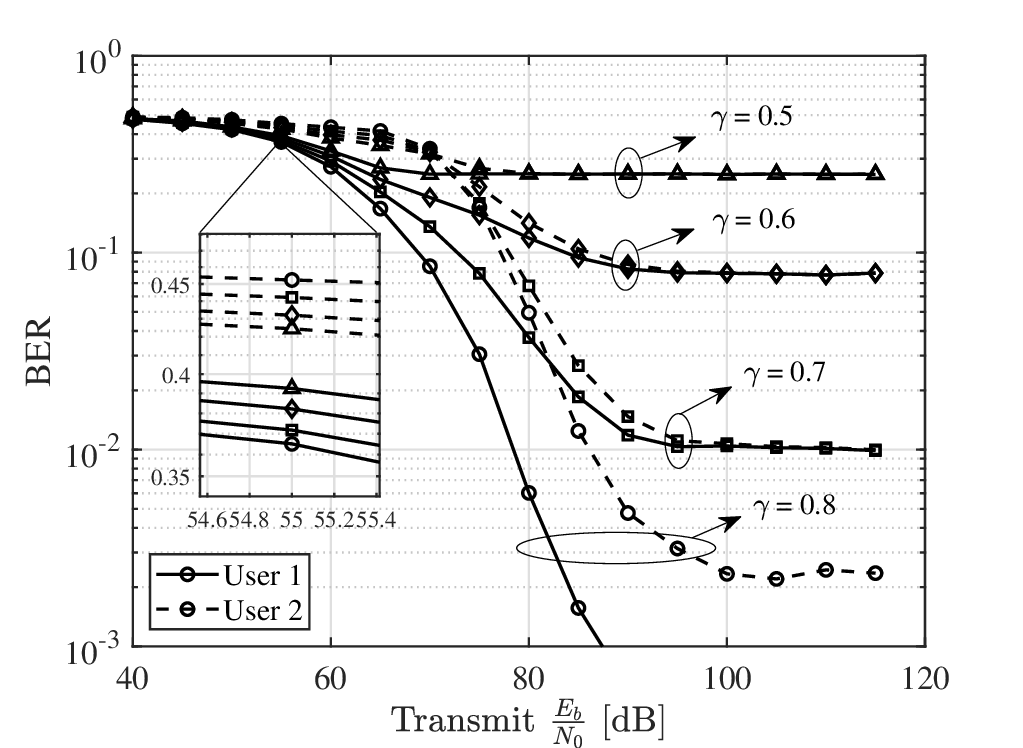}
    \caption{BER performance of the proposed system.}
    \label{fig:BER}
\end{figure}
\begin{figure}[ht]
    \centering
    \includegraphics[scale = 0.45]{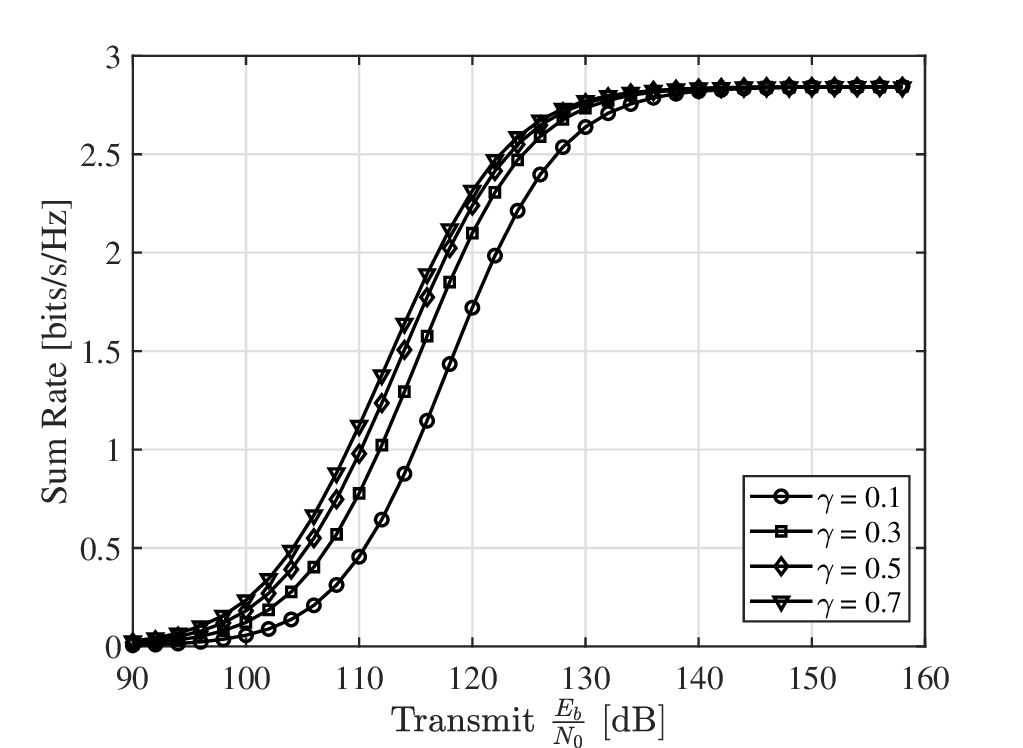}
    \caption{Sum-rate performance of the proposed system.}
    \label{fig:sumrate}
\end{figure}
\begin{figure}[ht]
    \begin{subfigure}[b]{0.49\linewidth}
     \centering
         \includegraphics[width= 1.3\linewidth, height =1.2\linewidth]{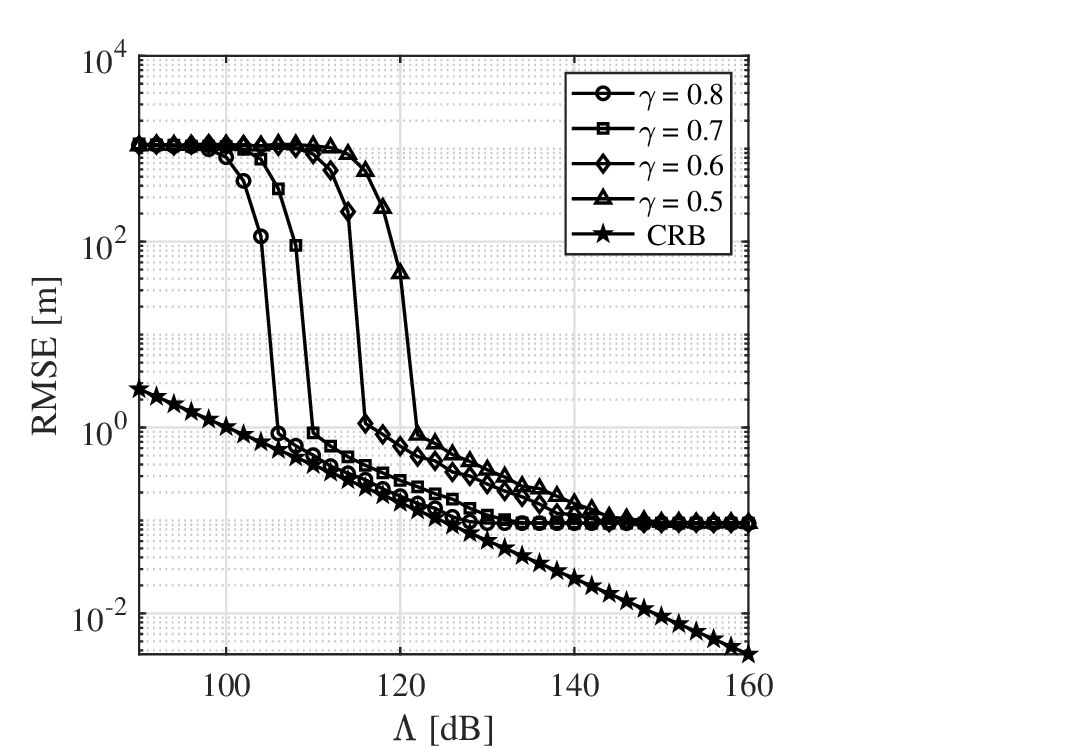}
         \caption{User 1.}
         \label{fig:RMSE_1}
     \end{subfigure}
     \hspace{-1em}
     \begin{subfigure}{0.49\linewidth}
     \centering
     \includegraphics[width= 1.3\linewidth, height=1.2\linewidth]{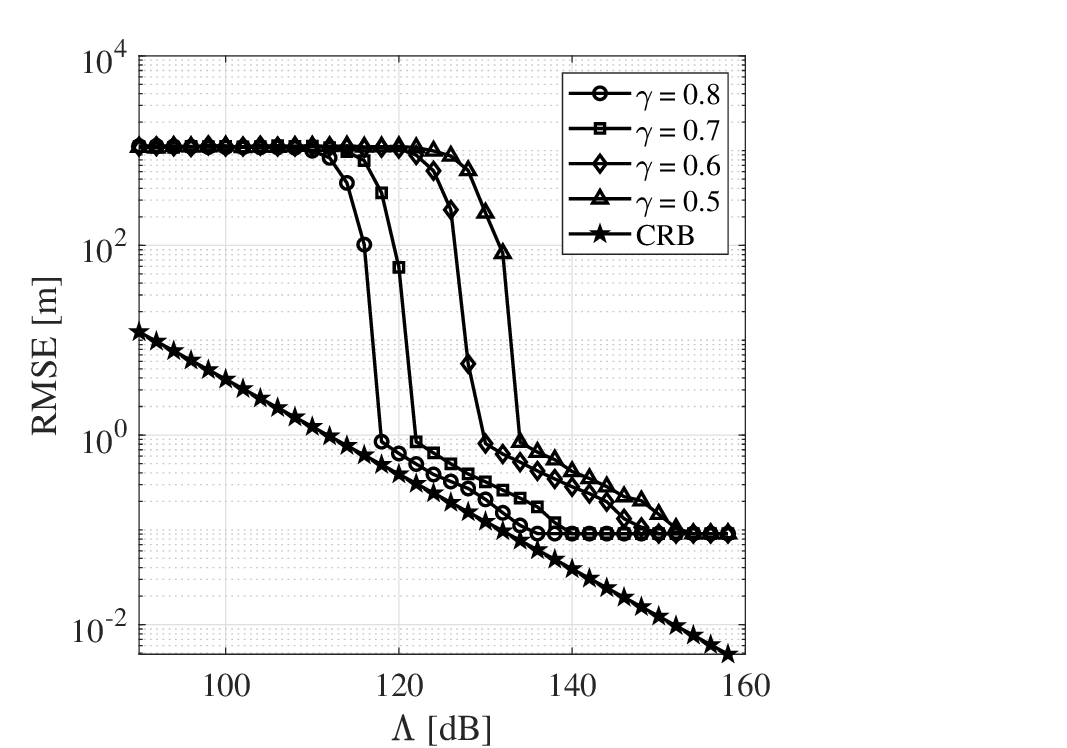}
      \caption{User 2.}
      \label{fig:RMSE_2}
     \end{subfigure}
     \caption{RMSE and CRB of the proposed system.}
\end{figure}

Fig.~5 illustrates the sum-rate performance of the proposed system for different power allocation factors $\gamma$ with respect to $\frac{E_b}{N_0}$. As shown, the sum-rate initially increases with $\frac{E_b}{N_0}$, reaches a peak, and then saturates. The reason for this behavior is similar to the case of BER. At low $\frac{E_b}{N_0}$, where clipping distortion is negligible, increasing $\frac{E_b}{N_0}$ improves the sum-rate. However, once $\frac{E_b}{N_0}$ exceeds a certain threshold (i.e., around 140 dB), clipping distortion becomes significant, and further increases in $\frac{E_b}{N_0}$ no longer enhance performance, causing the sum-rate to saturate at $2.8$ bits/s/Hz.
Moreover, the results indicate that a more balanced power allocation (i.e., smaller $\gamma$) leads to a lower overall sum-rate, whereas allocating more power to the stronger user (i.e., larger $\gamma$) enhances the total performance, as the user with better channel conditions dominates the sum-rate.

The RMSE and square root CRB for link distance estimations of Users 1 and 2 are plotted with respect to~$\Lambda$ in Figs.~\ref{fig:RMSE_1} and~\ref{fig:RMSE_2}, respectively. It is seen that the RMSE saturates at different~$\Lambda$ values depending on the user’s channel and power allocation~$\gamma$. When power is equally allocated (i.e., $\gamma = 0.5$), the RMSE decreases most slowly with $\Lambda$, whereas allocating more power to User~1 results in faster reduction. This occurs because equal power allocation produces no dominant component in the NOMA signal, causing random phase fluctuations that impair peak detection after the FFT. For instance, for $\gamma = 0.5$ and $0.8$, the RMSE in the case of User~1 reaches the saturation of $9.13~\mathrm{cm}$ at $\Lambda = 146$ dB and $130$ dB, respectively. In the case of User~2, although the RMSE shows a similar trend, it decreases more slowly due to weaker channel gain. Ultimately, both users converge to the same RMSE floor of $9.13~\mathrm{cm}$, which is fundamentally limited by system bandwidth.

\section{Conclusion}
This paper proposed a multi-user ISAC system for VLC employing NOMA and DCO-OFDM, with particular attention to the impact of clipping distortion. By applying clipping before NOMA superposition coding, the proposed system was able to reduce the PAPR compared to existing schemes. Simulation results showed that while allocating more power to the stronger user improves both the sum rate and sensing performance, a more balanced power distribution across users tends to reduce sensing performance. These findings highlight the inherent trade-off between fairness in power allocation and the joint optimization of communication and sensing. Future work will focus on experimental validation and the exploration of adaptive power allocation and clipping strategies in more dynamic scenarios.
\bibliographystyle{ieeetr}   
\bibliography{references}
\end{document}